\documentclass[twocolumn, twocolappendix]{aastex631}

\newcommand{\zn}{J0923$+$0402}
\newcommand{\kms}{km s$^{-1}$}
\newcommand{\lbol}{$L_{\rm Bol}$}

\shorttitle{Feedback in a Lo-BAL quasar at Reionization}
\shortauthors{Bischetti, Choi, et al.}
\graphicspath{{./}{images/}}
\usepackage{CJKutf8}
\begin{document}

\begin{CJK}{UTF8}{}
\CJKfamily{mj}

\title{Multi-phase black-hole feedback and a bright [CII] halo in a Lo-BAL quasar at $\mathbf{z\sim6.6}$}

\correspondingauthor{M. Bischetti, H. Choi}
\email{manuela.bischetti@units.it, hyunseop.choi@umontreal.ca}

\author[0000-0002-4314-021X]{Manuela Bischetti}
\affiliation{Dipartimento di Fisica, Universit\'a di Trieste, Sezione di Astronomia, 
Via G.B. Tiepolo 11, I-34131 Trieste, Italy}
\affiliation{INAF - Osservatorio Astronomico di Trieste, Via G. B. Tiepolo 11, I–34131 Trieste, Italy}

\author[0000-0002-3173-1098]{Hyunseop Choi (최현섭)}
\affiliation{D\'epartement de Physique, Universit\'e de Montr\'eal, Succ. Centre-Ville, Montr\'eal, QC, H3C 3J7, Canada}
\affiliation{Mila - Quebec Artificial Intelligence Institute, Montr\'eal, QC, Canada}

\author[0000-0002-4031-4157]{Fabrizio Fiore}\affiliation{INAF - Osservatorio Astronomico di Trieste, Via G. B. Tiepolo 11, I–34131 Trieste, Italy}
\affiliation{IFPU - Institut for fundamental physics of the Universe, Via Beirut 2, 34014 Trieste, Italy }

\author[0000-0002-4227-6035]{Chiara Feruglio}\affiliation{INAF - Osservatorio Astronomico di Trieste, Via G. B. Tiepolo 11, I–34131 Trieste, Italy}
\affiliation{IFPU - Institut for fundamental physics of the Universe, Via Beirut 2, 34014 Trieste, Italy }

\author[0000-0002-6719-380X]{Stefano Carniani}\affiliation{Scuola Normale Superiore, Piazza dei Cavalieri 7, I-56126 Pisa, Italy}

\author[0000-0003-3693-3091]{Valentina D'Odorico}
\affiliation{INAF - Osservatorio Astronomico di Trieste, Via G. B. Tiepolo 11, I–34131 Trieste, Italy}
\affiliation{IFPU - Institut for fundamental physics of the Universe, Via Beirut 2, 34014 Trieste, Italy }
\affiliation{Scuola Normale Superiore, Piazza dei Cavalieri 7, I-56126 Pisa, Italy}

\author[0000-0002-2931-7824]{Eduardo Ba\~nados}
\affiliation{Max Planck Institut für Astronomie, Königstuhl 17, D-69117, Heidelberg, Germany}

\author[0000-0002-3211-9642]{Huanqing Chen}
\affiliation{Canadian Institute for Theoretical Astrophysics, University of Toronto, 60 St George St, Toronto, ON M5R 2M8, Canada}

\author[0000-0002-2662-8803]{Roberto Decarli}
\affiliation{INAF - Osservatorio di Astrofisica e Scienza dello Spazio di Bologna, via Gobetti 93/3, I-40129, Bologna, Italy}

\author[0000-0002-7200-8293]{Simona Gallerani}\affiliation{Scuola Normale Superiore, Piazza dei Cavalieri 7, I-56126 Pisa, Italy}

\author[0000-0001-7271-7340]{Julie Hlavacek-Larrondo}
\affiliation{D\'epartement de Physique, Universit\'e de Montr\'eal, Succ. Centre-Ville, Montr\'eal, QC, H3C 3J7, Canada}

\author[0000-0001-9372-4611]{Samuel Lai}
\affiliation{Research School of Astronomy and Astrophysics, Australian National University, Canberra, ACT 2611, Australia
}

\author[0000-0002-3809-0051]{Karen M.\ Leighly}
\affiliation{Homer L.\ Dodge Department of Physics and Astronomy, The
  University of Oklahoma, 440 W.\ Brooks St., Norman, OK 73019}

\author[0000-0002-5941-5214]{Chiara Mazzucchelli}
\affiliation{N\'ucleo de Astronom\'ia de la Facultad de Ingenier\'ia, Universidad Diego Portales, Av. Ej\'ercito Libertador 441, Santiago, Chile}

\author[0000-0003-3544-3939]{Laurence Perreault-Levasseur}
\affiliation{D\'epartement de Physique, Universit\'e de Montr\'eal, Succ. Centre-Ville, Montr\'eal, QC, H3C 3J7, Canada}
\affiliation{Mila - Quebec Artificial Intelligence Institute, Montr\'eal, QC, Canada}
\affiliation{Ciela - Montreal Institute for Astrophysical Data Analysis and Machine Learning, Montréal, Canada}
\affiliation{Center for Computational Astrophysics, Flatiron Institute, 162 5th Avenue, 10010, New York, NY, USA}
\author[0000-0002-9909-3491]{Roberta Tripodi}
\affiliation{University of Ljubljana FMF, Jadranska 19, 1000, Slovenia}
\affiliation{INAF - Osservatorio Astronomico di Trieste, Via G. B. Tiepolo 11, I–34131 Trieste, Italy}
\affiliation{IFPU - Institut for fundamental physics of the Universe, Via Beirut 2, 34014 Trieste, Italy }

\author[0000-0001-5287-4242]{Fabian Walter}
\affiliation{Max Planck Institut für Astronomie, Königstuhl 17, D-69117, Heidelberg, Germany}

\author[0000-0002-7633-431X]{Feige Wang}
\affiliation{Steward Observatory, University of Arizona, 933 N Cherry Avenue, Tucson, AZ 85721, USA}

\author[0000-0001-5287-4242]{Jinyi Yang}
\affiliation{Steward Observatory, University of Arizona, 933 N Cherry Avenue, Tucson, AZ 85721, USA}

\author[0000-0001-7883-496X]{Maria Vittoria Zanchettin}
\affiliation{SISSA, Via Bonomea 265, 34136, Trieste, Italy}
\affiliation{INAF - Osservatorio Astronomico di Trieste, Via G. B. Tiepolo 11, I–34131 Trieste, Italy}

\author[0000-0003-3307-7525]{Yongda Zhu}
\affiliation{Department of Physics \& Astronomy, University of California, Riverside, CA 92521, USA}



\begin{abstract}
   Although the mass growth of supermassive black holes during the Epoch of Reionisation is expected to play a role in shaping the concurrent growth of their host-galaxies, observational evidence of feedback at z$\gtrsim$6 is still sparse. We perform the first multi-scale and multi-phase characterisation of black-hole driven outflows in the $z\sim6.6$ quasar J0923+0402 and assess how these winds impact the cold gas reservoir. We employ the SimBAL spectral synthesis to fit broad absorption line (BAL) features and find a powerful ionized outflow on $\lesssim210$ pc scale, with a kinetic power $\sim2-100$\% of the quasar luminosity. ALMA observations of [CII] emission allow us to study the morphology and kinematics of the cold gas. We detect high-velocity [CII] emission, likely associated with a cold neutral outflow at $\sim0.5-2$ kpc scale in the host-galaxy, and a bright extended [CII] halo with a size of $\sim15$ kpc. For the first time at such an early epoch, we accurately constrain the outflow energetics in both the ionized and the atomic neutral gas phases. We find such energetics to be consistent with expectations for an efficient feedback mechanism, and both ejective and preventative feedback modes are likely at play. The scales and energetics of the ionized and atomic outflows suggest that they might be associated with different quasar accretion episodes. The results of this work indicate that strong black hole feedback is occurring in quasars at $z\gtrsim6$ and is likely responsible for shaping the properties of the cold gas reservoir up to circum-galactic scales.
\end{abstract}



\section{Introduction} \label{sec:intro}

Bright quasars at $z\gtrsim6$ are powered by billion solar mass black-holes (BHs) that lie well above the local BH mass - galaxy dynamical mass correlation \citep{Neeleman21}, implying that BH growth dominates over galaxy growth during the first Gyr of the Universe \citep{Volonteri12,Inayoshi22,Hu22}. Cosmological simulations predict such a rapid BH growth to drive powerful winds at $z\sim6-7$ \citep{Barai18,vanderVlugt19}. These winds are responsible for suppressing further BH growth and regulating star formation activity and the physical properties of the interstellar (ISM) and circum-galactic (CGM) media, possibly leading to the BH and host-galaxy co-evolution observed at lower redshift \citep{Valentini21,Pizzati23, Costa22}. 
In the past years, a major effort was undertaken to detect BH-driven winds from emission lines in the millimetre band, resulting in only a few detections and conflicting results about the occurrence and strength of cold gas outflows at $z\gtrsim6$ \citep[e.g.][]{Maiolino12, Cicone15, Bischetti19, Novak20, Izumi21, Tripodi22, Meyer22}. 

Recent deep spectroscopic surveys of quasars at $z\gtrsim6$ showed that about half of the quasar population shows broad absorption line (BAL) features in the rest-frame UV spectrum, tracing ionized gas winds \citep{Schindler20,Bischetti22,Bischetti23,Yang21}, while at $z\lesssim4$ the BAL quasar fraction is 10-20\% \citep{Maddox08,Gibson09}. BAL winds can arise in the nuclear regions of galaxies and, in $\gtrsim50$\% of quasars, reach 0.1-10 kpc scale \citep{Arav18,Choi22}; hence they may represent a source of feedback on both BH and host-galaxy growth.
Outflow energetics have been measured with good accuracy in several BAL quasars at $z\lesssim3$, finding indeed that BALs provide an efficient feedback mechanism \citep{Fiore17,Choi20,Choi22,Miller20}. However, little is known about the gas physical properties and energetics of BALs in $z\gtrsim6$ quasars.

Here we report on \zn\ \citep[$z=6.626$, ][]{Bischetti22}, which is the highest redshift low-ionization broad absorption line (Lo-BAL) quasar discovered so far. This object was initially classified as a low-luminosity, dust-reddened quasar based on SUBARU/HSC and WISE photometry \citep{Matsuoka18b}, but revealed to be a brighter quasar with little extinction \citep{Wang19UKIRT, Kato20}. Indeed, \zn\ has a bolometric luminosity log($L_{\rm Bol}/\rm erg\ s^{-1})=47.5$ and a black-hole mass log($M_{\rm BH}/\rm M_\odot)=9.4$ \citep{Yang21,Mazzucchelli23}, which implies that it is accreting close to the Eddington limit ($L_{\rm Bol}/L_{\rm Edd}\sim1$, where $L_{\rm Edd}$ is the Eddington luminosity). The detection of strong C IV, Si IV, and Mg II absorption features associated with a BAL wind in \zn\ was reported in \cite{Bischetti22,Bischetti23}.
We exploit the \textit{SimBAL} \citep{Leighly18} spectral analysis to measure the energetics of the BAL wind in \zn. We also present ALMA observations of the [CII] $\lambda158\mu\rm m$ emission line to trace the galaxy-scale counterpart of the BAL wind and assess the impact of black-hole activity on ISM and CGM properties. Throughout the paper, we adopt $\Lambda$CDM cosmology with $H_0 = 67.3$ \kms, $\Omega_\Lambda = 0.69$, and $\Omega_M$ = 0.31 \citep{Planck16}.

\section{Data analysis and results}

\subsection{BAL data and \textit{SimBAL} analysis}\label{subsec:simbal}

We use the spectral-synthesis code \textit{SimBAL} \citep{Leighly18} to fit the BAL features in the rest-UV spectrum of \zn, which was acquired with the VLT/X-shooter spectrograph as part of the XQR-30 survey \citep{Dodorico23}. Details on the data reduction and calibration are given in \cite{Bischetti22, Dodorico23},
The spectrum was shifted to the rest frame using $z_{\rm [CII]}$  (Table \ref{table:almaprop}, calculated as in Sect. \ref{subsec:alma}) and binned to increase the signal-to-noise ratio (SNR$\gtrsim 5$ per $100\ \mathrm{km\ s^{-1}}$ pixel in the CIV trough at$\sim1500$ \AA).
We fit the rest-frame spectral range 1280-2900 \AA\, in which the main broad absorption features are observed.

\begin{figure*}
    \centering
    \includegraphics[width = 0.62 \linewidth]{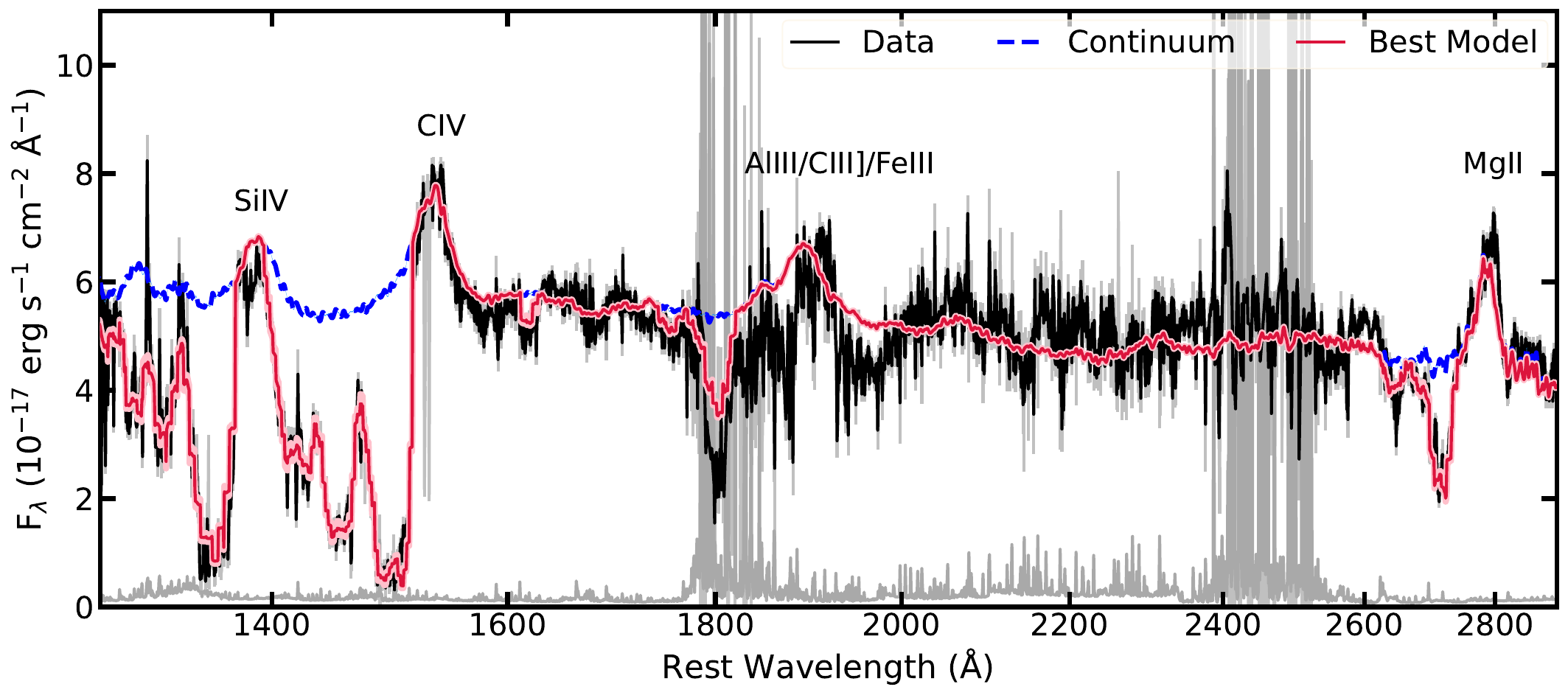}
    \includegraphics[width =0.37\linewidth]{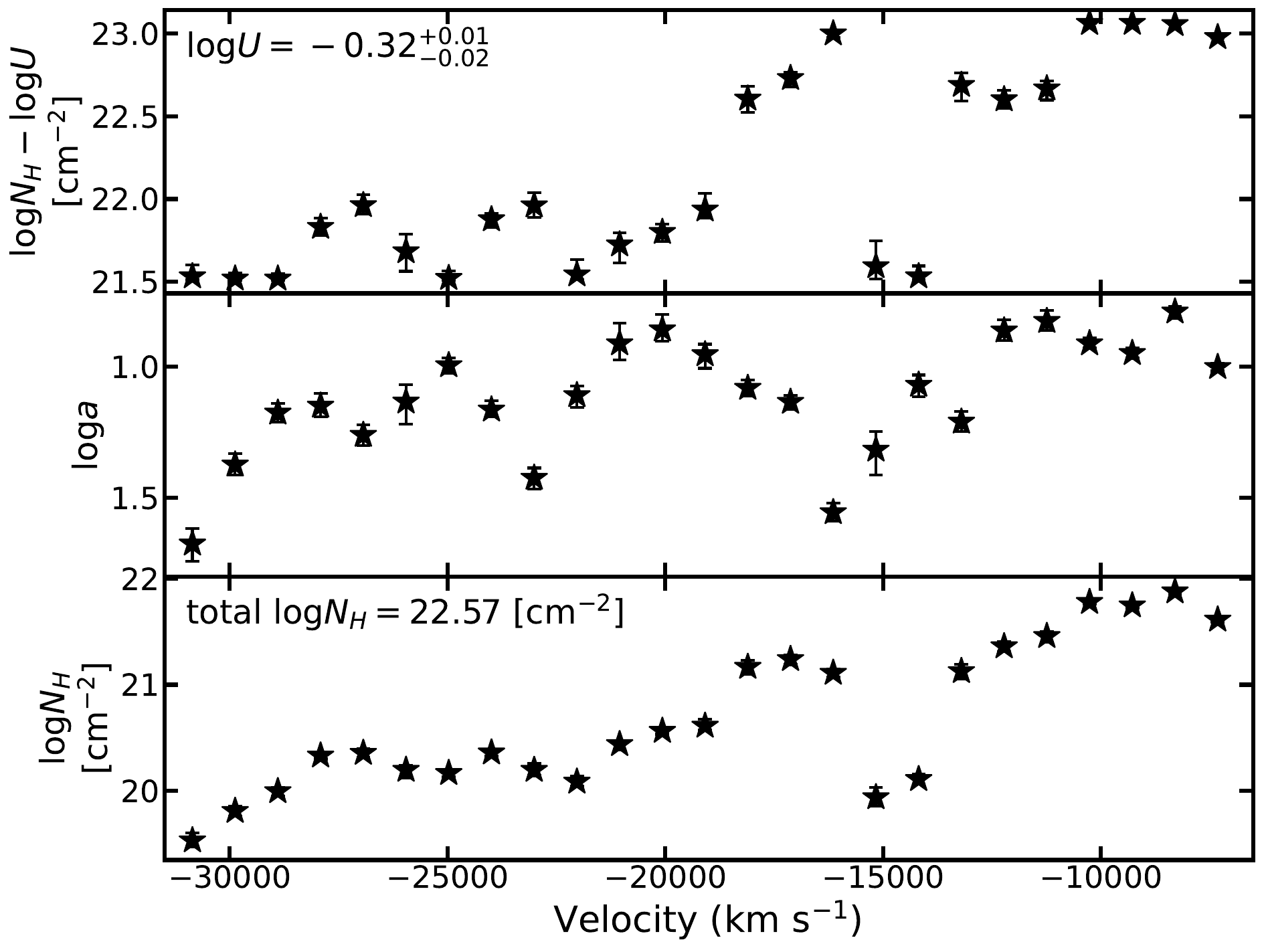}
    \caption{\textit{Left:} X-Shooter spectrum of \zn\ (black) and best-fitting \textit{SimBAL} model (red).
    The dashed blue curve represents the unabsorbed quasar emission model by \citep{Bischetti23}.
    The noise spectrum and the spectral regions excluded from the fit are plotted in grey.
    Labels indicate the major UV emission lines. \textit{Right:} BAL outflow physical parameters as a function of velocity with error bars representing 95\% confidence regions.
    The top two panels show the fit parameters ($\log\ U$, $\log N_H-\log U$, $\log\ a$) used in \textit{SimBAL} modeling.
    The bottom panel shows the distribution of hydrogen column density ($\log N_H$) corrected for the covering fraction (Appendix~\ref{App:simbal}).
    }
    \label{fig:simbal_bestfit}
\end{figure*}

We use a velocity-resolved model for the BAL absorption features \citep{Leighly18,Choi20} to fit the broad opacity profile observed in \zn.
We consider 25 bins with velocities spanning from $\sim-32,000\ \mathrm{km\ s^{-1}}$ to $\sim-6,800\ \mathrm{km\ s^{-1}}$\footnote{We retain the signs for the velocities as calculated from the quasar rest frame and compare their magnitudes \citep[i.e., $-10,000\ \mathrm{km\ s^{-1}}$ is a ``lower'' velocity than $-30,000\ \mathrm{km\ s^{-1}}$; e.g.,][]{Choi22b}}.
Each bin is $\sim 1,000$ \kms\ wide, which is sufficient to model the velocity profile of the troughs observed in the X-Shooter data while avoiding overfitting the BAL features.
The wind properties (e.g., column density, mass outflow rate) were calculated from the sum of the values calculated for each velocity bin.
The version of \textit{SimBAL} used in this work uses the grids calculated from version C17 of \textit{Cloudy} \citep{Ferland17}, which assumes solar metallicity and spectral energy distribution (SED) of an ionizing spectrum that is suitable for luminous quasars \citep[see][for a discussion of the \textit{SimBAL} updates]{Leighly18,Choi20,Choi22}.
To model the unabsorbed quasar emission, we used the composite template spectrum from \citet{Bischetti23}. We fitted the rest-frame spectral range 1280-2900 \AA\, in which the main broad absorption features are observed.

We consider a single ionization parameter ($\log\ U$), while
column density (parameterized as $\log N_H-\log U$), and partial covering parameter ($\log\ a$) were allowed to vary across different bins.
As the X-Shooter data show no BAL features associated with transitions sensitive to gas density \citep[e.g.,][]{Lucy14}, we did not include density as a free parameter in the \textit{SimBAL} modeling.
Instead, we fixed the density at log$(n/\rm cm^{-3})=6$ and verified that the best-fit parameters do not significantly change ($\lesssim0.15$ dex) considering the large range of densities $2.8\lesssim$log($n/\rm cm^{-3}$)$\lesssim8$ typically measured for LoBALs and FeLoBALs\footnote{LoBAL quasars that also show broad absorption associated with FeII transitions} \citep{Leighly18,Choi22}. These variations are indeed small compared to the ranges of outflow energetics we calculate in Sect. \ref{subsec:BAL_energy}.

The best-fit model is shown in Figure~\ref{fig:simbal_bestfit}.
Overall, we find an excellent agreement between the model and spectrum, the sole discernible deviation in AlIII absorption, which falls within a region heavily affected by strong tellurics.
We obtain a robust constraint of the ionization parameter of the BAL gas with $\log~U=-0.32^{+0.01}_{-0.02}$, thanks to 
the presence of BAL transitions from both low-ionization ions (e.g., AlIII, MgII) and high-ionization ions (e.g., CIV).
We find that the bulk of the column density is concentrated at the lower velocities (Fig. \ref{fig:simbal_bestfit} right).
This trend is reflected in the observed BAL structures, in which the LoBAL transitions are found only at the low-velocity end ($v_{out}\gtrsim-13,000\ \mathrm{km\ s^{-1}}$, Figure~\ref{fig:simbal_ions}).
The total hydrogen column density of the outflow $\log (N_H/\rm cm^{-2})=22.57^{+0.01}_{-0.02}$ was derived by correcting the column densities for partial covering (Appendix~\ref{App:simbal}) and summing over all bins.

The covering fraction parameter $\log a$ also varies greatly with velocity.
The shape of the CIV trough roughly matches the velocity structure of $\log a$ \citep[low covering corresponds to high values of $a$; see][]{Leighly19}.
This is because the line profile of a saturated BAL is highly dependent on the properties of partial covering \citep[e.g.,][]{deKool02}.
Our best-fitting model suggests that the BAL profile is produced by a combination of change in column density and partial covering across the velocity rather than by gas ionization structure, given a single-ionization parameter model is sufficient to recreate the profile.
Uncertainties were calculated from the posterior probability distribution of model parameters.
Further details on the \textit{SimBAL} modeling of \zn\ can be found in Appendix~\ref{App:simbal}.

\subsection{ALMA data}\label{subsec:alma}

\begin{figure*}
    \centering
    \includegraphics[width = 0.34\linewidth]{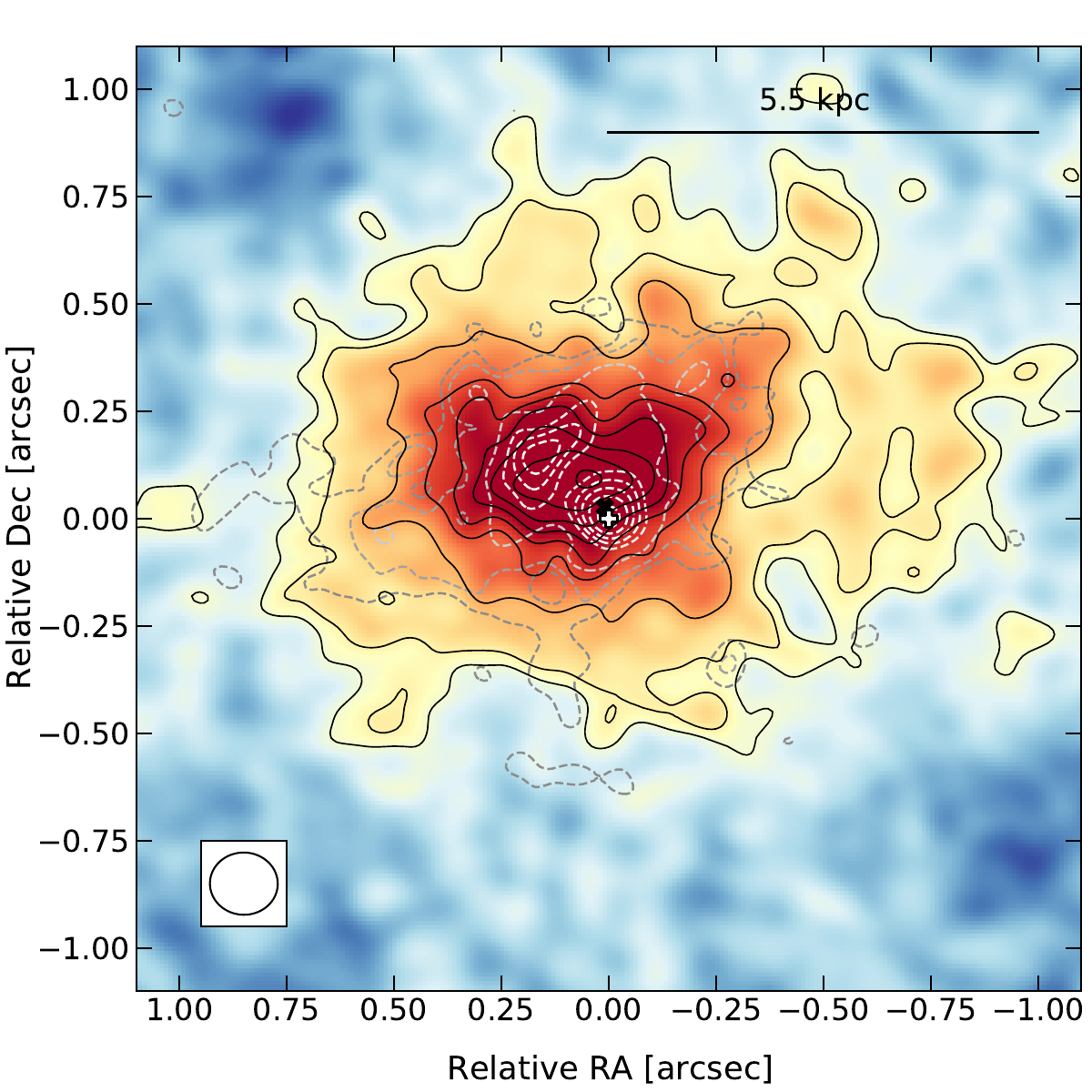}
    \includegraphics[width = 0.34\linewidth]{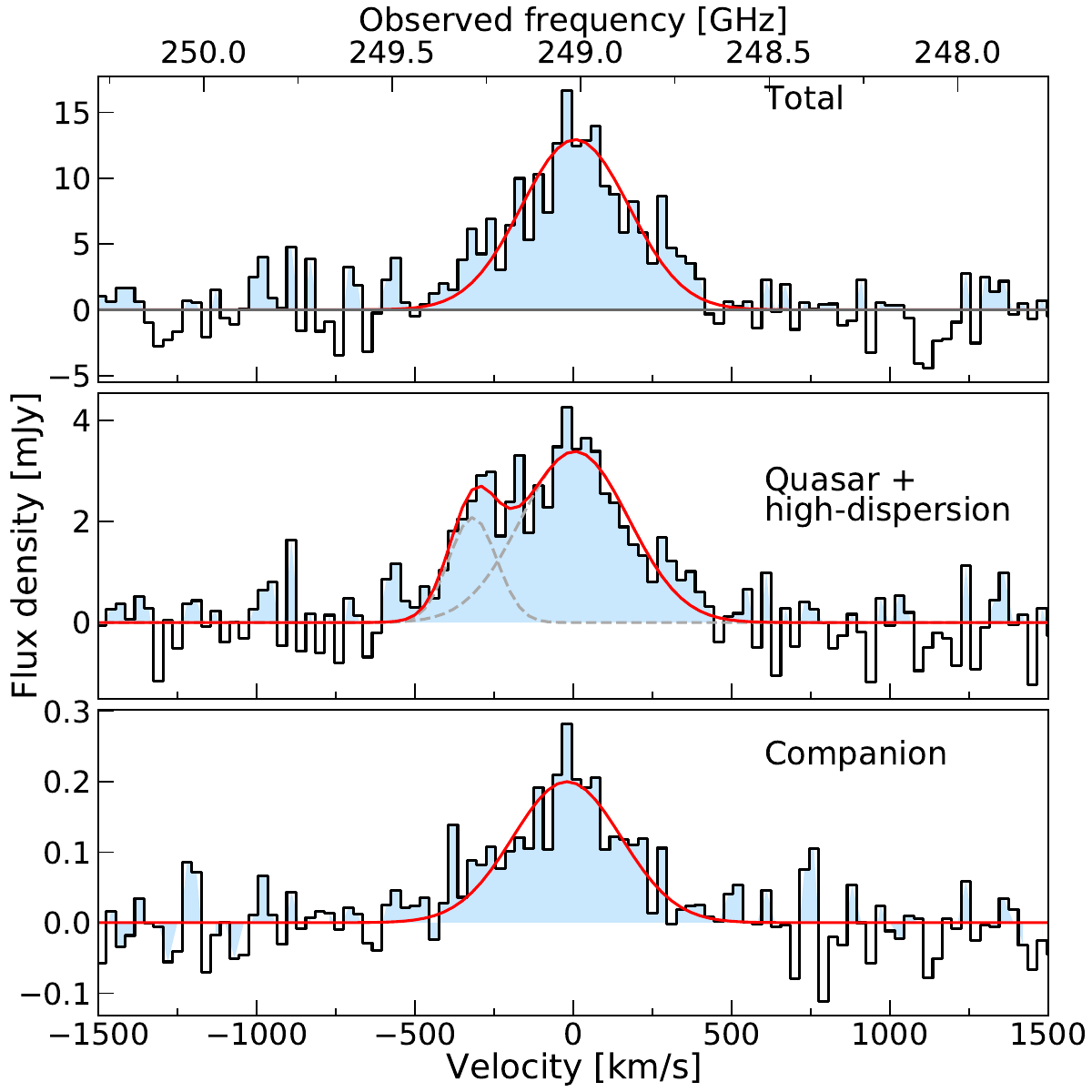}
    \includegraphics[width = 0.28\linewidth]{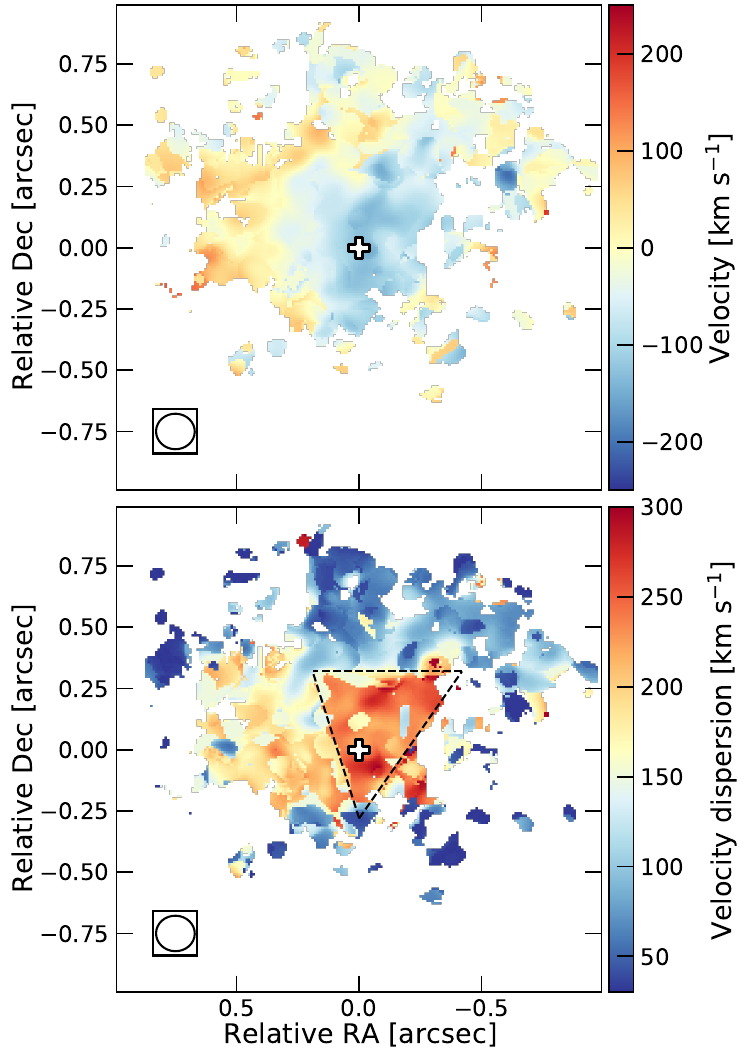}
    \caption{{\it Left:} Map of the [CII] emission in \zn. Solid contours correspond to [3,4,6,8,10,12,13]$\sigma$ significance, where $\sigma=0.017$ Jy beam$^{-1}$ \kms. The same levels are shown for the continuum emission by the dashed contours, with $\sigma=5.9\times10^{-6}$ Jy beam$^{-1}$. The black cross indicates the quasar optical position, as measured from NIRCam F200W data, and the white plus sign corresponds to the peak of the 242-257 GHz continuum emission probed by ALMA. \textit{Middle:} [CII] spectrum corresponding to the total emission detected at $\gtrsim3\sigma$ (top panel). The best-fit Gaussian profile is shown by the red curve. Middle panel displays the [CII] spectrum extracted from the region with velocity dispersion higher than 230 \kms, including the quasar location. To model the line profile, we added a second blueshifted Gaussian component with $v\simeq-300$ \kms\ to the scaled best-fit model of the top panel. Bottom panel shows the [CII] spectrum extracted from one beam aperture centred on the secondary continuum peak. {\it Right:} Velocity and velocity dispersion maps corresponding to the [CII] emission at $>3\sigma$. The black-dashed polygon highlights the triangle-shaped region with high-velocity dispersion (Sect. \ref{subsec:alma}).}
    \label{fig:cii-integrated}
\end{figure*}

We analyse archival ALMA observations of \zn\ from projects 2018.1.01188S, 2019.1.00111S, and 2021.1.00934S. These datasets target [CII]($\nu_{\rm rest}=1900.5369$ GHz) and band 6 continuum emission (observed frequency $\sim242-257$ GHz) with angular resolution varying from $\sim$0.12 to 0.57 arcsec. 
Visibilities were calibrated using the standard calibration provided by the ALMA observatory and the default phase, bandpass and flux calibrators. To maximise the sensitivity and, at the same time, improve the sampling of the uv-plane, we merged the visibilities from the three datasets.  We created continuum maps by averaging visibilities over all four 1.875 GHz spectral windows, excluding the spectral range covered by [CII] emission. To model and subtract the continuum emission to the line, we combined the adjacent spectral windows in the baseband containing [CII] and performed a fit in the $uv$ plane to channels with $|v|>750$ \kms, using a first-order polynomial continuum. Continuum-subtracted datacubes were created using CASA task $tclean$, with the $hogbom$ cleaning algorithm in non-interactive mode, a threshold equal to two times the rms sensitivity and a natural weighting of the visibilities. For [CII] datacubes, we adopted a 30 \kms\ channel width. The resulting synthesized beam is $0.16"\times0.15"$ for the spectral window including [CII] (rms sensitivity $7.8\times10^{-5}$ Jy beam$^{-1}$ for a 30 \kms\ channel width), and $0.14"\times0.13"$ for the continuum image (rms sensitivity $5.9\times10^{-6}$ Jy beam$^{-1}$). 

We detect a bright [CII] emission, with a velocity integrated flux density of $S_{\rm [CII]}\simeq6.1$ Jy \kms, corresponding to a luminosity $L_{\rm [CII]}\simeq6.7\times10^9$ L$_\odot$. The [CII] spectrum extracted from the 3$\sigma$ isophote contours is shown in Fig. \ref{fig:cii-integrated} middle. The peak of the [CII] emission is located at an observed frequency of 249.019 GHz, corresponding to a $z_{[\rm CII]}=6.6321$, which is similar to the [CII]-based redshift previously reported by \cite{Yang21} (the difference is less than one channel of our ALMA datacube). Fitting a 2D Gaussian to the [CII] map gives a FWHM size of ($1.28\times0.92$) arcsec$^2$, that is about ($7.1\times5.1$) kpc$^2$. 
We also detect the $\sim242-257$ GHz continuum emission (Fig. \ref{fig:cii-integrated} right) with a flux density  $S_{\rm cont}\simeq0.8$ mJy (Table \ref{table:almaprop}). This corresponds to an infrared luminosity $L_{\rm IR,8-1000\mu m}\simeq2.4\times10^{12}$ L$_{\odot}$, assuming a typical dust temperature of 47 K and a dust emissivity index $\beta=1.6$ of high-redshift quasar host galaxies \citep{Beelen06, Tripodi23}.
The continuum morphology shows a main peak close to the quasar optical position, based on JWST NIRCam observations (F200W filter, project GO 2078, black cross in Fig. \ref{fig:cii-integrated}). We measure a $0.13"$ offset with respect to the quasar position from SUBARU Hyper Suprime-Cam (HSC) observations \citep{Kato20}, and do not find evidence for astrometric errors in the HSC image, although we caution that the size of an HSC pixel is $0.17"$. We also detect a second continuum peak eastern of the main one , separated by $\simeq1.2$ kpc, possibly indicating a companion galaxy. Due to the presence of diffuse emission between and around these continuum peaks, it is not straightforward to derive a continuum flux ratio between the quasar host galaxy and the companion. We provide a basic estimate by independently fitting a 2D gaussian profile to the two continuum peaks, and find a quasar-to-companion flux ratio of $\sim2.5:1$. The maximum of the [CII] emission is located between these two continuum peaks, about $\simeq550$ pc offset from the main one.  

Figure \ref{fig:cii-integrated} right shows the velocity and velocity dispersion maps of the [CII] emission \zn. We do not find a strong velocity gradient close to the quasar location, as the bulk of [CII] shows projected velocities within $\pm100$ \kms. {This suggests that the quasar host-galaxy is seen under relatively low inclination, as typically expected for BAL quasars \citep[e.g.][]{Elvis00}}. Moderately redshifted emission is detected $\sim3$ kpc east from the nucleus, in a region in which we also find extended continuum emission, likely due to a tidal feature. We also find blueshifted velocities at the location of the quasar and in a triangle-shaped region, extending up to 0.4" from it. Such velocities are due to a second peak in the [CII] spectrum, blueshifted by $v\simeq-300$ \kms\ with respect to the systemic emission and reaching $v\simeq-500$ \kms\ (Fig. \ref{fig:cii-integrated} middle). The [CII] velocity dispersion is generally high ($\sigma_v\simeq150-200$ \kms) in the central 1.5-2 kpc, the highest values ($\simeq 300$ \kms) being also located in the triangle-shaped region. The combination of high-velocity and high-velocity dispersion suggests the presence of a [CII] outflow in the host-galaxy of \zn. Due to the moderate velocities observed, we cannot a priori exclude that the triangle-shaped region is associated with a tidal feature. However, as high-velocity ($v<-250$ \kms) [CII] emission is very faint at the location of the companion galaxy (Fig. \ref{fig:cii-integrated} middle), and is instead mostly located north-west from the quasar (Fig. \ref{fig:cii-integrated} right, and Fig. \ref{fig:chmaps}), we deem it unlikely that this emission is due to the merger traced by the continuum emission.  The peak of this blueshifted emission is located about $0.1"$ (520 pc) from the nucleus, and its flux is about 7\% of the total [CII] flux (Table \ref{table:almaprop}), similar to what previously measured for other [CII] outflows at $z\gtrsim6$ \citep[e.g.][]{Bischetti19,Izumi21, Izumi21a}.

\begin{table}
    \setlength{\tabcolsep}{2pt}
    \caption{Properties of \zn.}
    \label{table:almaprop}
    \centering
    \begin{tabular}{lc}
    
    \hline\hline
    
    \hline
    RA & 09:23:47.122\\
    Dec & +04:02:54.402\\
    $z_{\rm [CII]}$ & 6.6321$\pm$0.0003 \\
    $S_{\rm cont}$ [mJy] & 0.80$\pm$0.08 \\
    $S_{\rm [CII]}$ [Jy \kms]    & 5.94$\pm$0.35\\
    FWHM$_{\rm [CII]}$ [\kms]    & 400$\pm$25\\
    L'$_{\rm [CII]}$ [10$^9$L$_\odot$] & 6.75$\pm$0.45 \\

    \hline
    \multicolumn{2}{c}{BAL outflow}\\
    $v_{\rm max}$ [\kms] &  -32,000 \\
    $\log\ U$ & $-0.32^{+0.01}_{-0.02}$\\
    $\log N_H-\log U$$^a$ [$\mathrm{cm^{-2}}$] & $21.52-23.06$\\
    $\log\ a$$^a$ & $0.79-1.67$ \\
    log($N_{\rm H} /[\rm cm^{-2}]$)$^b$ & $22.57^{+0.01}_{-0.02}$ \\
    $R_{\rm BAL}$ [$\mathrm{pc}$] & $3-210$ \\
    log($\dot{M}_{\rm BAL}/[\rm M_\odot\ yr^{-1}]$) & $1.89-3.7$\\
    log($\dot{E}_{\rm BAL}/[\rm erg\ s^{-1}]$)   & $45.69-47.51$\\
    log($\dot{P}_{\rm BAL}/[\rm g\ cm\ s^{-2}]$) & $36.81-38.62$\\
    \hline

    \multicolumn{2}{c}{[CII] outflow}\\
    
    $S_{\rm [CII]}$ [Jy \kms]    & 0.40$\pm$0.07\\    
    $v_{\rm [CII]}$ [\kms] & -530$\pm$30  \\
    $R_{\rm [CII]}$   [pc] & $520-2,220$\\
    log($\dot{M}_{\rm atom}/[\rm M_\odot\ yr^{-1}]$) & $2.18-2.81$\\
    log($\dot{E}_{\rm atom}/[\rm erg\ s^{-1}]$) & $43.11-43.76$\\
    log($\dot{P}_{\rm atom}/[\rm g\ cm\ s^{-2}]$) & $35.85-36.51$ \\

    \hline
    \end{tabular}
    
    \tablecomments{RA, Dec refer to the peak of the 242-257 GHz continuum. Uncertainties correspond to a 68\% confidence level for the [CII] outflow properties and to 95\% for those of the BAL outflow. The range of outflow radii and energetics correspond to the plausible range of locations of the outflowing gas (\S~\ref{subsec:BAL_energy}). $^a$ The range of values extracted from the multiple velocity bins is reported. 
    $^b$ Corrected for the covering fraction (Appendix~\ref{App:simbal}).}
    
\end{table}

\section{Discussion and Conclusions} 
\begin{figure*}
    \centering
    \includegraphics[width = 1\textwidth]{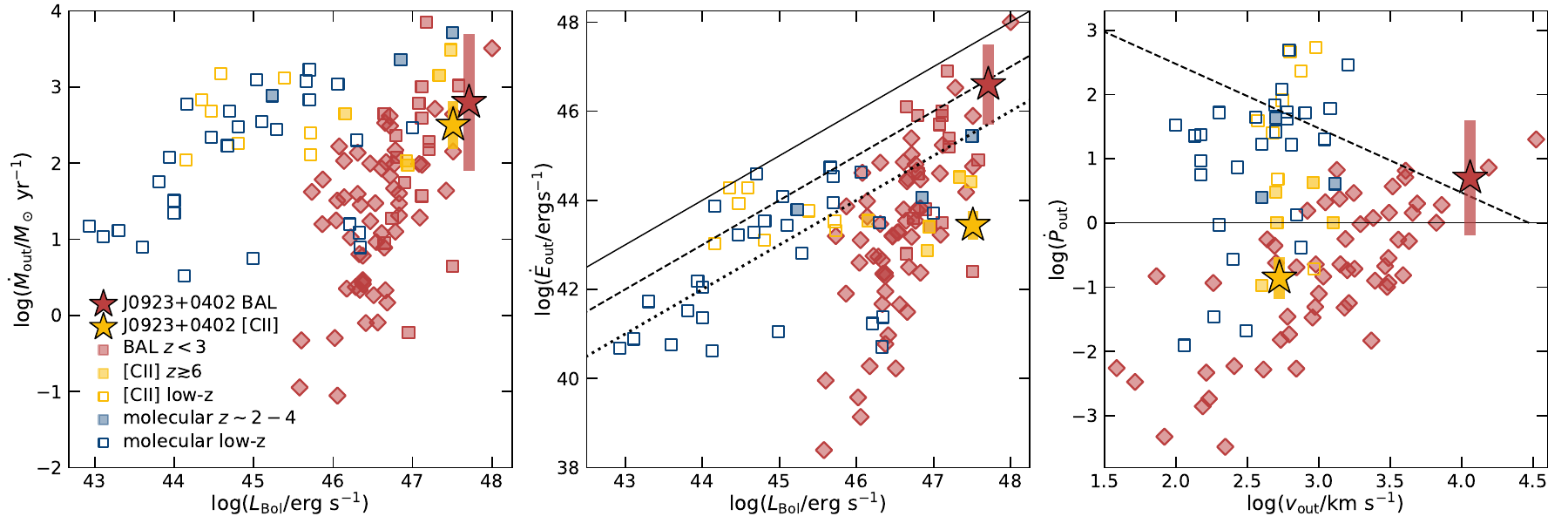}
    \caption{{\it Left:} Mass outflow rate as a function of bolometric luminosity for the [CII] and BAL outflow detected in \zn\ (stars). The shaded bars show the range of possible values according to our analysis. We compare our results with [CII] outflows from \cite{Bischetti19, Izumi21a, Izumi21, Tripodi22} and BAL outflows from \cite{Choi20,Choi22} (diamonds) and \cite{Miller20} (squares). We also include the compilation of black-hole driven molecular outflows from \cite{Bischetti19pds} and references therein, and recent measurements by \cite{RamosAlmeida22, Zanchettin21, Zanchettin23}.
    The estimates for the comparison BAL outflows were also calculated assuming a thin-shell scenario.
    The bolometric luminosity for the BAL outflow of \zn\ is arbitrarily shifted for plotting. {\it Middle:} Outflow kinetic power as a function of $L_{\rm Bol}$. The solid, dashed and dotted lines correspond to $L_{\rm Bol}$, 0.1$L_{\rm Bol}$, and 0.01$L_{\rm Bol}$, respectively. {\it Right:} Outflow momentum flux as a function of the outflow velocity. The solid(dashed) line indicates the expected momentum boost for a black-hole driven wind arising with nuclear velocity 0.1$c$ and propagating with a momentum(energy) conserving expansion \citep{Zubovas&King12}. }
    \label{fig:correlations}
\end{figure*}

\subsection{Wind location and energetics \label{subsec:BAL_energy}}

We determine the plausible range of BAL wind radii $3\lesssim R_{\mathrm{BAL}}\lesssim 210$ pc based on the assumption that the BAL outflow energy does not exceed \lbol\/ (upper limit) and by assuming that the density of the BAL absorbing gas is lower to that of gas located in the quasar broad line region (BLR), consistently with previous measurements of BAL physical properties and energetics from $z<3$ quasars  (Appendix~\ref{App:bal_radii}).
We calculate the mass outflow rate from the best-fit $N_H$ and $v$ using
Eq. (9) in \citet{Dunn10}, which adopts a thin-shell geometry,
and assuming a mean molecular weight $\mu=1.4$, a global covering $\Omega=0.2$ \citep[e.g.,][]{Hewett03}.
The latter is derived from the fraction of BAL quasars at $z\lesssim4$. For $z\sim6$ quasars, the frequency increases to almost 50\% \citep{Bischetti22}, which may suggest a global covering fraction as high as 0.5,
which would yield a higher estimate of the mass outflow rate.
Based on the range of BAL wind radii,
we measure a mass outflow rate $1.9\lesssim\log(\dot M_{\rm BAL}/\rm M_\odot\ yr^{-1})\lesssim3.7$ (Table \ref{table:almaprop}), which is is among the largest values measured for BAL winds in quasars at $z\lesssim3$ \citep[][Fig. \ref{fig:correlations} left]{Choi20,Choi22,Miller20, Bruni19, Vietri22}.
Similarly, we calculated the total mass of the BAL wind to be in the range $4.0 \lesssim$ log($M_{\rm BAL}/\rm M_\odot$)$\lesssim7.7$, considering a shell-like scenario.
The kinetic power and momentum flux of the BAL wind are listed in Table \ref{table:almaprop}.

For the putative [CII] outflow in \zn, we calculate an atomic gas mass $M_{\rm atom}\simeq4.3\times10^8$ $M_\odot$ using Eq. (1) in \cite{Hailey-Dunsheath10} and typical assumptions on the physical conditions of the [CII] emitting gas. We assume a C$^+$ abundance of $1.4\times10^{-4}$ per hydrogen atom and a gas temperature of 200 K \citep{Kaufman99, Hailey-Dunsheath10}. Indeed, a temperature of a few 100 K is expected in the outflow even in the molecular gas phase \citep{Richings18}. A temperature in the range 100-1000 K would imply a variation of only 20\% in $M_{\rm atom}$. As we do not know the gas density in the [CII] outflow, we consider the density limit $n>>n_{\rm crit}$, where $n_{\rm crit} = 2.7\times10^3$ cm$^{-3}$ is the [CII] critical density, to derive a lower limit to the outflowing gas mass \citep{Bischetti19}. This limit is consistent with the large density ($n\simeq10^{5}$ cm$^{-3}$) measured in the host-galaxy of the $z\sim6$ quasar J1148$+$5251 by \cite{Maiolino05} and with the large densities found in several quasar outflows \citep{Aalto12,Aalto15}. A lower gas density \citep[e.g.][]{Meyer22} might be expected for a kpc-scale outflow and would correspond to a higher mass of the [CII] outflow. 

We compute the [CII] mass outflow rate using Eq. (1) in \cite{Bischetti19pds} for a shell-like outflow geometry \citep{Lutz20, Izumi21a, Izumi21}. To account for the uncertainty on the [CII] outflow location, we either consider the maximum extent or the scale of the bulk of the high-velocity [CII] emission (\S~ \ref{subsec:alma}). This results in $\dot{M}_{\rm atom}\simeq150-640$ $M_\odot$ yr$^{-1}$, similar to what has been previously reported for [CII] outflows detected at $z\gtrsim6$ in a few individual quasars or via stacking analysis \citep{Maiolino12,Cicone14,Izumi21a,Izumi21,Tripodi22,Bischetti19}, and in low-redshift active galactic nuclei with {\it Herschel}/SPIRE observations \citep{Janssen16}. We calculate the kinetic power of the BAL and [CII] outflows as $\dot{E} = \dot{M}v^2/2$ and the outflow momentum load $\dot{P}_{\rm} = \dot{M}v/(L_{\rm Bol}/c)$, that is the outflow momentum flux normalised to the quasar radiation momentum flux (Table \ref{table:almaprop}). 

\subsection{Multiphase and multiscale outflow}

\zn\ shows evidence of a relatively compact ($\lesssim210$ pc) BAL wind of ionised gas and of a neutral atomic outflow extending up to galactic scale ($520-2200$ pc). The [CII] outflow is more extended than the BAL, but if we consider the upper limit on $R_{\rm BAL}$, their sizes are in order-of-magnitude agreement. 
Most of the outflowing gas mass is carried by the neutral outflow ($10\lesssim M_{\rm atom}/M_{\rm BAL}\lesssim4\times10^4$), similar to the findings for low-redshift active galactic nuclei \citep{Fluetsch19, Speranza23}. The mass outflow rates for the two outflow phases are roughly similar (Fig. \ref{fig:correlations} left), while the bulk of the outflow luminosity is associated with the ionised phase, as the BAL kinetic power is consistent with a significant, if not a major, fraction of the quasar luminosity ($0.02\lesssim \dot{E}_{\rm BAL}/L_{\rm Bol}\lesssim1.0$, Fig. \ref{fig:correlations} middle). Such a high luminosity matches the expectations for an efficient quasar feedback mechanism \citep[e.g.][]{Faucher12,Costa14}, and supports the scenario previously suggested by \cite{Bischetti22,Bischetti23} in which BAL quasars witness strong quasar feedback occurring at $z\gtrsim6$. The mass outflow rate of the BAL wind is larger than the mass accretion rate necessary to power the observed quasar luminosity, assuming a standard 10\% accretion efficiency \citep[$57\ M_\odot\ \mathrm{yr^{-1}}$,][]{Marconi04}. This implies that quasar feedback will likely suppress further black-hole growth.
Despite hosting such a powerful BAL, \zn\ shows at most a mild perturbation of the cold gas kinematics in its central region, as only modest [CII] velocities are observed. A possible explanation for this might be that the BAL imprint on the ISM kinematics is diluted by the modest resolution ($\sim800$ pc) of the ALMA observations. This may imply a small coupling of the ionized gas in the outflow with the [CII] gas, especially if the bulk of [CII] is coming from photo-dissociated regions (PDRs). It could also be interpreted as the BAL having a small impact on the ISM because of the relatively low opening angle of the wind \citep[e.g.][]{Menci19,Bischetti19}.

Concerning possible driving mechanisms of the outflows in \zn, we find $\dot{P}<1$ for the atomic phase (Fig. \ref{fig:correlations} right), while models of quasar outflows based on the energy-driven expansion of a radiatively-driven nuclear wind predict load factors $\gg10$ \citep{Faucher12, Zubovas&King12}. Differently, if the BAL wind is located on scales $\gtrsim10$ pc, its momentum load is large enough to be in agreement with an energy-conserving expansion. This difference between the neutral and ionized outflows in \zn\ suggests that the two outflows might not be associated with the same accretion episode.  The flow timescales ($\tau\sim R/v$) of the outflows are also very different,  $\tau_{\rm BAL}\lesssim1.4\times10^4$ yrs and $\tau_{\rm atom}\simeq(1-4)\times10^6$ yrs, respectively. While the BAL is likely powered by the observed \lbol, the [CII] outflow might be a remnant from a previous accretion episode, similar to the fossil outflows discovered in low-redshift active galactic nuclei \citep{Fluetsch19,Bischetti19pds,Zubovas23}. Multiple episodes of strong quasar feedback, spaced out by phases of gas accretion due to mergers or inflows from the intergalactic medium, are indeed expected by cosmological simulations of high-redshift quasars \citep[e.g.][]{vanderVlugt19,Zubovas21}.

From FWHM$_{\rm [CII]}$  (Table \ref{table:almaprop}) we can derive a rough estimate of the dynamical mass using Eq. (14) by \cite{Neeleman21} and, in turn, of the escape velocity within the inner $\sim2$ kpc, that is $v_{\rm esc}\simeq500$ \kms. As the velocity of the [CII] outflow is similar to $v_{\rm esc}$, a significant fraction of the outflowing gas might be able to reach CGM scales in a few Myr.
The feedback mechanism occurring in \zn\ thus likely involves both channels of the quasar feedback paradigm, preventative and ejective \citep[e.g.][]{Somerville15,CurtisLake23}. While the neutral outflow can displace or remove gas from the host-galaxy, the ionised wind injects large amounts of energy into the ISM and CGM, thereby reducing the accretion of gas at later epochs \citep{Tumlinson17}.

\begin{figure}
    \centering
    \includegraphics[width = 1\columnwidth]{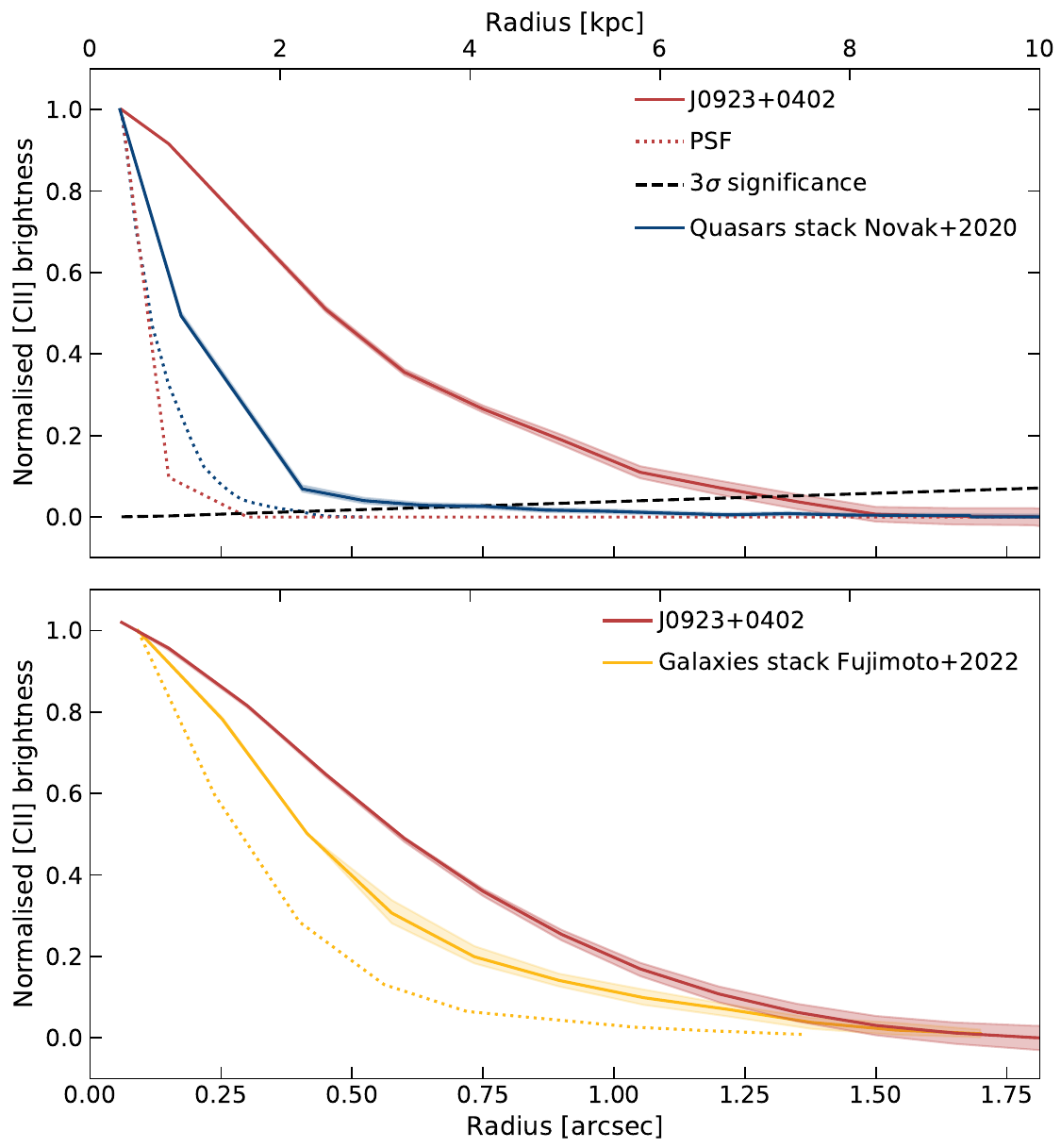}
    \caption{{\it Top:} Brightness profile of the [CII] emission in \zn\ (red solid curve), and normalised to the maximum value. The brightness profile associated with the ALMA point spread function (PSF) is shown by the dotted red curve, while the dashed curve shows the profile associated with the $3\sigma$ level. 
    The [CII] stacked profile of $z\sim6$ quasar hosts,  including mergers, by \citep{Novak20} is shown for comparison (PSF$\sim0.18"$). Shaded areas indicate 68\% confidence level uncertainties. {\it Bottom:} Comparison of the [CII] brightness profile of \zn\ and the stacked profile of massive star forming galaxies by \cite{Fujimoto19}. We convolved our data to the  PSF$\sim0.4"$ by \cite{Fujimoto19} to compare sizes for a similar angular resolution.}
    
    \label{fig:profile}
\end{figure}

\subsection{Bright [CII] halo: is there a link with feedback?}
ALMA observations of \zn\ reveal the presence of a bright [CII] emission associated with the host galaxy ISM and CGM. 
[CII] mostly traces neutral atomic gas in PDRs and is often used as a tracer of star-formation in high-redshift galaxies \citep{Maiolino05,Decarli18}, although it can be also emitted from the partly ionised medium \citep{Carilli13,Lagache18}.
In the case of \zn, the [CII] luminosity is $\sim0.6$ dex brighter than what is typically observed in quasars with similar $L_{\rm IR,8-1000\mu m}$ \citep{Decarli18, Venemans20}, suggesting that
a significant fraction of its [CII] emission may not arise from PDRs but may rather trace diffuse and ionised gas in the CGM. This is supported by the very different morphology of continuum and [CII] emission (Fig. \ref{fig:cii-integrated}), and by about 50\% of the [CII] emission arising beyond a radius of $\simeq3.3$ kpc (Fig. \ref{fig:profile}). 

[CII] emission on scales beyond a few kpc has been previously detected by stacking samples of star forming galaxies at $z\gtrsim5$, and in a few individual targets \citep{Fujimoto19,Fujimoto20, Ginolfi20, Herrera-Camus21, Akins22, Lambert23}. 
Systematic studies in $z\gtrsim6$ quasars have instead found that typical [CII] radii are $<2$ kpc \citep{Novak20, Venemans20}, and that the upper envelope of these values is often associated with recent/ongoing mergers \citep{Neeleman21}. Radii $>3$ kpc have been reported in only few cases, based on moderate resolution observations ($0.5-1"$)\citep{Cicone15,Izumi21a}. We calculate the brightness profile of [CII] emission in \zn\ in circular annuli of $0.15"$ radius, following the method described in \cite{Tripodi22}, and find that size of the [CII] halo around \zn\ is four times more extended than the stacked [CII] profile by \cite{Novak20}, and it is also larger than the stacked halo of massive galaxies obtained by \cite{Fujimoto19} after removing possible active galaxies in their sample. 

Although \zn\ likely has a merging companion {located $\sim1$ kpc eastward} from the nucleus, the morphology of the [CII] emission is mostly symmetric (the east-west versus north-south size ratio is $\sim1.2$, considering the $3\sigma$ contours in Fig. \ref{fig:cii-integrated} left) and extends to much larger radii. This is different from what is typically observed via [CII] in merging systems, that is a [CII] bridge connecting quasar and companion, instead than a diffuse halo \citep[e.g.][]{Tripodi24, Decarli19}. The fact that mergers typically modify the [CII] morphology along a specific direction is also supported by the fact that \cite{Novak20} do not find a difference in the [CII] stacked profiles of $z\sim6$ quasars with/without mergers. The above picture questions the merger as the main mechanism responsible for the extended [CII] halo \citep[e.g.,][]{Ginolfi20b, Lambert23}.
An alternative scenario concerns [CII] emission being powered when streams of dense and cold gas accrete on high-redshift galaxies and are heated due to gravitational energy release and shock heating \citep[e.g.][]{Dekel09}. However, in such a scenario, the metallicity of the inflowing gas is expected to be low and not to significantly contribute to [CII] emission \citep{Pallottini14, Vallini15}.

Recent works have suggested a link between feedback and extended gaseous halos around $z\sim6$ star forming galaxies. Given the high luminosity and the evidence of multiple outflow episodes in \zn, such a scenario might explain its extended [CII] emission.  On one hand, feedback is able to displace gas beyond few kpc \citep{Costa19,Vito22} and, at the same time, outflows can significantly contribute to gas heating via shocks \citep{Appleton13,Fujimoto20, Pizzati23}. If this is the case for \zn, the [CII] halo could be a remnant of past feedback activity \citep{Pizzati23}, as we do not currently observe emission from high-velocity gas on the CGM scale.
On the other hand, black-hole feedback can regulate the fraction of photons reaching the scales of the CGM \citep{Costa22}. Along the lines of sight with high escape fraction (e.g., those cleared out by the BAL wind), the halo would be mostly ionized \citep[e.g.,][]{Obreja24}, consistently with the Ly$\alpha$ halos frequently detected around $z\gtrsim6$ quasars \citep{Farina19}. If the global escape fraction remains relatively low \citep{Stern21}, extended and bright [CII] halos are expected in massive halos such as those of $z\gtrsim6$ quasars \citep{Pizzati23, Costa23}.

The high redshift of \zn\ implies that Ly$\alpha$ emission lies at the very edge of the MUSE spectral band, thus severely challenging the detection of a Ly$\alpha$ halo around this quasar. Cycle 3 JWST observations with NIRSpec IFU will allow us to investigate the possible presence of companion galaxies non detected by ALMA, map the morphology and kinematics of the warm ionized gas phase in the host-galaxy and up to the CGM of \zn, as traced by the H$\alpha$ and [OIII] emission lines \citep[e.g.,][]{Leibler18, Marshall23}.


\begin{acknowledgments}
We are grateful to the anonymous referee for useful feedback which helped us to improve the paper. We thank Kastitis Zubovas and Tiago Costa for valuable suggestions during the preparation of this manuscript.
This paper makes use of the following ALMA data: ADS/JAO.ALMA\#2018.1.01188S (P.I. F. Wang), ADS/JAO.ALMA\#2019.1.00111S (P.I. B. Venemans), and ADS/JAO.ALMA\#2021.1.00934S (P.I. J, Yang). ALMA is a partnership of ESO (representing its member states), NSF (USA), and NINS (Japan), together with NRC (Canada), MOST and ASIAA (Taiwan), and KASI (Republic of Korea), in cooperation with the Republic of Chile. The Joint ALMA Observatory is operated by ESO, AUI/NRAO and NAOJ. This work is based in part on observations made with the NASA/ESA/CSA James Webb Space Telescope. 
The JWST data presented in this article were obtained from the Mikulski Archive for Space Telescopes (MAST) at the Space Telescope Science Institute. The specific observations analyzed can be accessed via \dataset[DOI: 10.17909/npbb-0x08]{https://doi.org/10.17909/npbb-0x08}.
These observations are associated with program GO 2078 (P.I. F. Wang). The project leading to this publication has received support from ORP, that is funded by the European Union’s Horizon 2020 research and innovation programme under grant agreement No 101004719 [ORP]. M.B. acknowledges support from INAF project 1.05.12.04.01 - MINI-GRANTS di RSN1 "Mini-feedback" and from UniTs under FVG LR 2/2011 project D55-microgrants23 "Hyper-gal". M.B., C.F. and F.F. acknowledge support from INAF PRIN 2022 2022TKPB2P - BIG-z.
Some of the computing for this project was performed at the OU Supercomputing Center for Education \& Research (OSCER) at the University of Oklahoma (OU).
JHL acknowledges the Canada Research Chair program, the NSERC discovery grant and the Discovery Accelerator Supplements program.
\end{acknowledgments}

%

\vspace{5mm}
\facilities{ALMA, VLT(X-Shooter), SUBARU(HSC), JWST(NIRCam)}.


\software{astropy \citep{Astropy13,Astropy18,Astropy22},  
          Cloudy \citep{Ferland17},
          SimBAL \citep{Leighly18}
          }



\clearpage
\appendix

\section{\textit{SimBAL} Analysis}\label{App:simbal_whole}

Before performing the \textit{SimBAL} analysis, we identified the data points and the regions in the VLT/X-shooter data to be excluded in the spectral modeling.
We masked the pixels heavily affected by telluric features by identifying any data points with an SNR$<3$.
Deep narrow intervening absorption lines at $\lambda\sim1550$ \AA, identified as MgII$\lambda\lambda 2796,2803$ at $z_{abs}=3.163$ \citep{Davies23}, were also masked.

\subsection{Best-fit \textit{SimBAL} model}\label{App:simbal}
\textit{SimBAL} fits both the continuum and absorption simultaneously to obtain a robust solution \citep[e.g.,][]{Choi20,Choi22}.
\textit{SimBAL} requires six physical parameters to create synthetic BAL features for a given outflow: the dimensionless ionization parameter $\log U$, the gas density $\log n\ \rm[cm^{-3}]$, a column density parameter $\log N_H-\log U\ \rm[cm^{-2}]$ which represents a hydrogen column density normalised by ionization parameter, the velocity offset $v_{off}\rm\ (km\ s^{-1})$, width of the absorption lines $v_{width}\rm\ (km\ s^{-1})$, and a dimensionless covering fraction parameter $\log a$.
The inhomogeneous partial covering model is adopted in \textit{SimBAL}, in which a power-law distribution of opacity \citep[$\tau =\tau_{max}x^a$ and $x\,\in\,(0,1)$,][] {sabra05,arav05} is controlled by $\log a$.
The full covering is achieved with low values of $a$ close to 0, and larger $a$ (or $\log a$) represents the low partial covering of the emission source.
Further discussion of inhomogeneous partial covering is given in \citet{Leighly19}.

We used \textit{SimBAL} version introduced in \citet{Choi20}, which uses an ionic column density grid calculated for a solar metallicity gas.
We expect a higher-metallicity grid would result in a smaller column density roughly inversely proportional to the metal enhancement factor and, therefore, weaker outflow strengths.
For instance, \citet{Leighly18} found in their analysis of a LoBAL SDSS~J0850+4451 that a 3 times higher metallicity grid would result in a smaller column density by a factor of about 3.
We used a set of 25 velocity-adjacent tophat bins \citep[tophat accordion model,][]{Leighly18} to model the absorption features.
The number of bins has a negligible impact on the result of the \textit{SimBAL} analysis unless it is set too low or high \citep{Leighly18}.
The bin width of $\sim 1,000\ \mathrm{km\ s^{-1}}$ proved to be sufficient to fit the BAL velocity structure observed in the binned X-Shooter data that we used for the \textit{SimBAL} analysis.

\begin{figure}[htb]
    \centering
    \includegraphics[width =1\columnwidth]{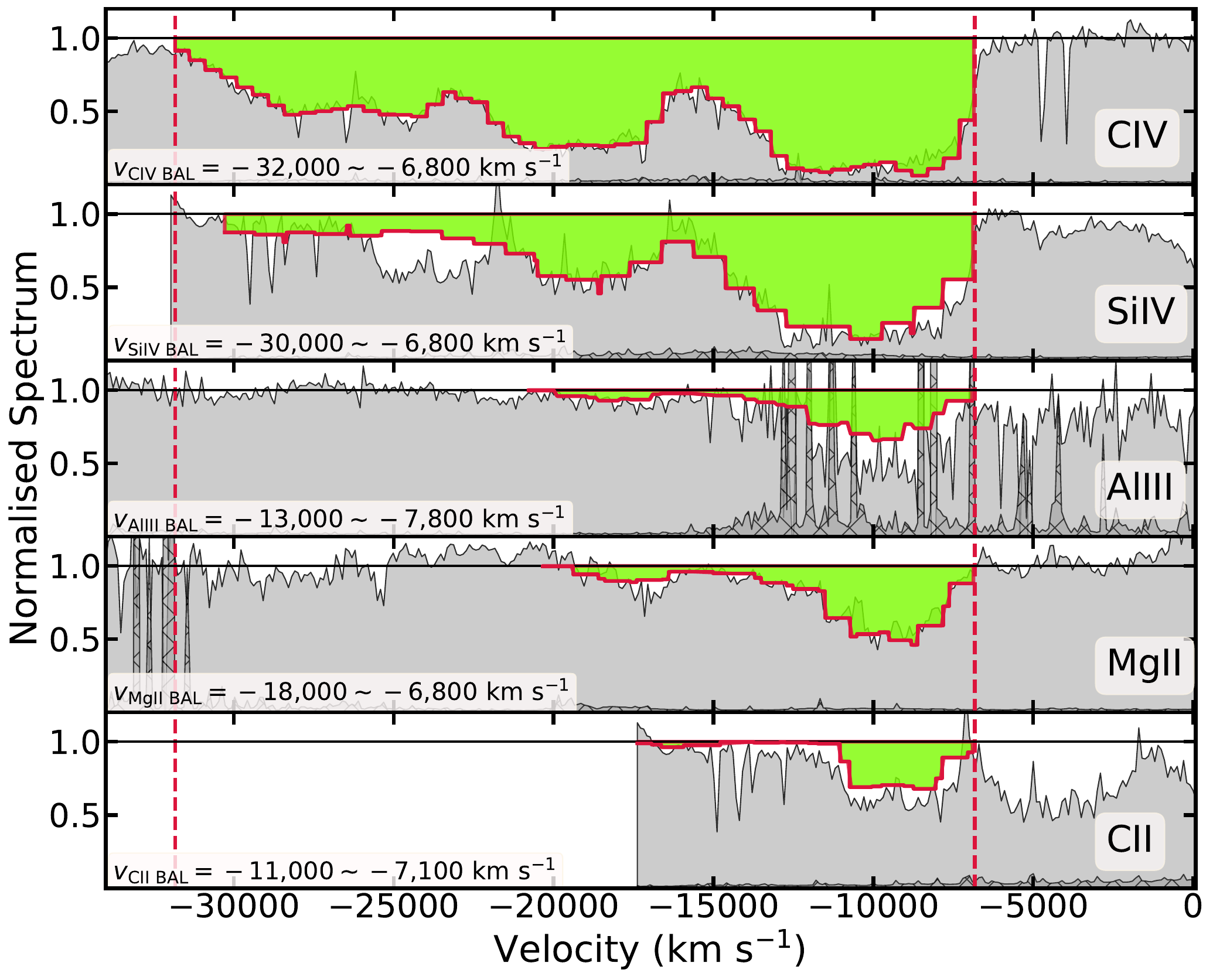}
    \caption{The normalised spectra of the X-Shooter data (grey shaded areas) and \textit{SimBAL} model (red curve, green shaded areas) plotted in the quasar rest-frame velocity.
    The dashed vertical lines highlight the maximum and minimum velocities of the BAL outflow.
    The CIV BAL represents the widest trough, and the LoBALs show significantly smaller velocity widths compared to the HiBAL troughs.
    We found that both HiBALs and LoBALs have similar minimum outflow velocities.
    The hatched grey area indicates the uncertainties associated with the data.
    \label{fig:simbal_ions}}
\end{figure}

\begin{figure}[htb]
    \centering
    \includegraphics[width =1\columnwidth]{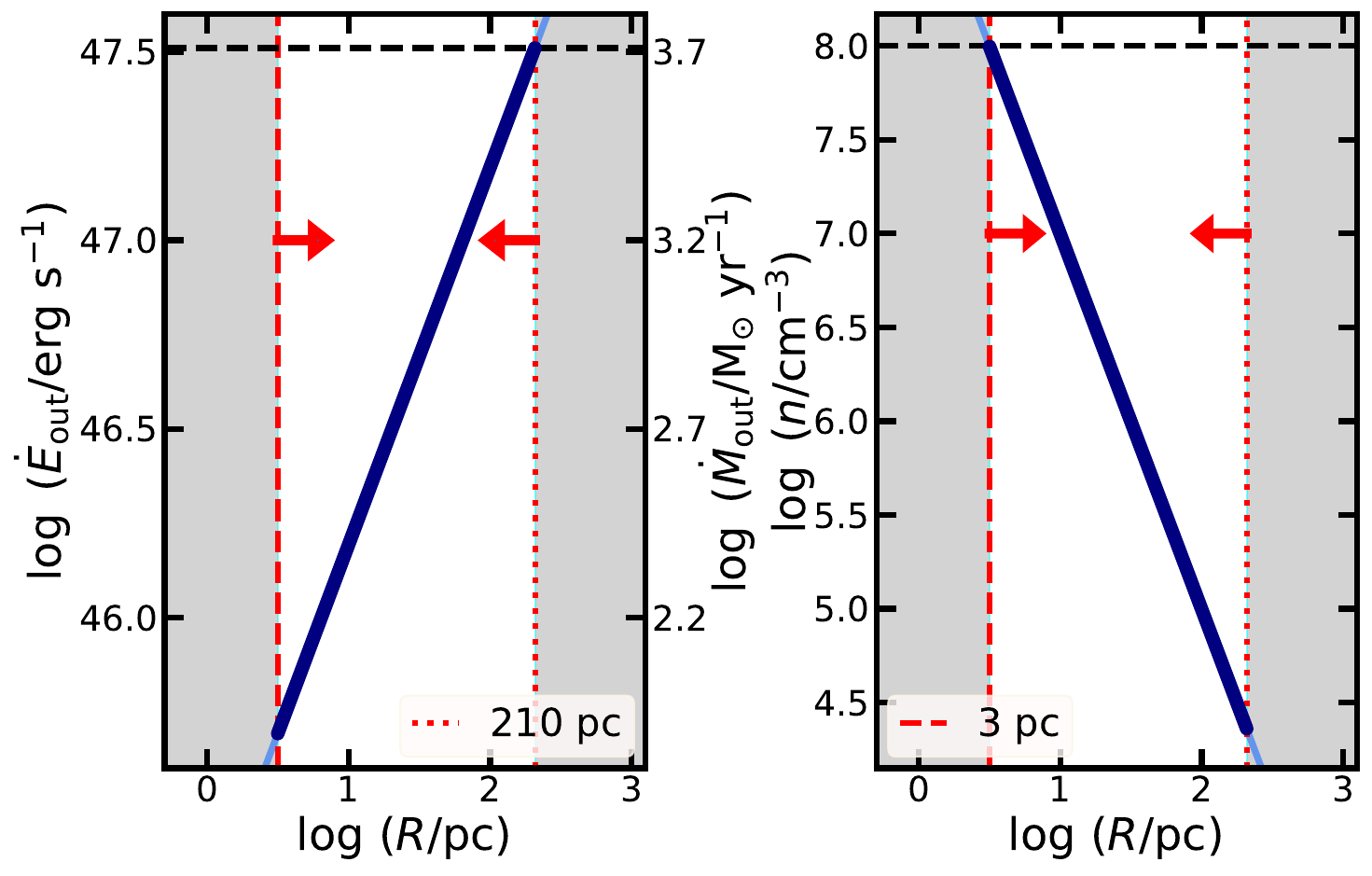}
    \caption{The range of BAL radii estimated from our analysis and the corresponding range of outflow properties.
    The dashed horizontal black lines represent the upper limits on the wind kinetic luminosity ($\dot E_{\mathrm{out}}\lesssim$\lbol) and the gas density, log($n/\rm cm^{-3}$)$\lesssim 8$, that provided us with both the upper and lower limits on the radius of BAL outflow (vertical red dotted and dashed lines, respectively).
    \label{fig:simbal_radii}}
\end{figure}

We extracted velocity-resolved physical properties from the best-fit \textit{SimBAL} model (Figure~\ref{fig:simbal_bestfit}) as well as the total column density of the outflowing gas to calculate the wind energetics (\S~\ref{subsec:BAL_energy}).
The best-fitting model presented in this work assumes a constant ionization parameter across the velocities.
We found that the data did not require a less restrictive model with multiple velocity-resolved ionization parameters.
In order to calculate the total column density, we first calculated column density values corrected for the covering fraction from each bin and summed the values \citep[$\log N_H=(\log N_H-\log U)+\log U-\log(1+10^{\log a})$,][]{arav05,Leighly18,Choi20,Green23}.
Similarly, the wind energetics, such as mass outflow rate, momentum flux, and kinetic luminosity, for the BAL were computed from the sum of the values calculated for each tophat bin.

\begin{figure*}[thb]
    \centering
    \includegraphics[width=0.7\textwidth]{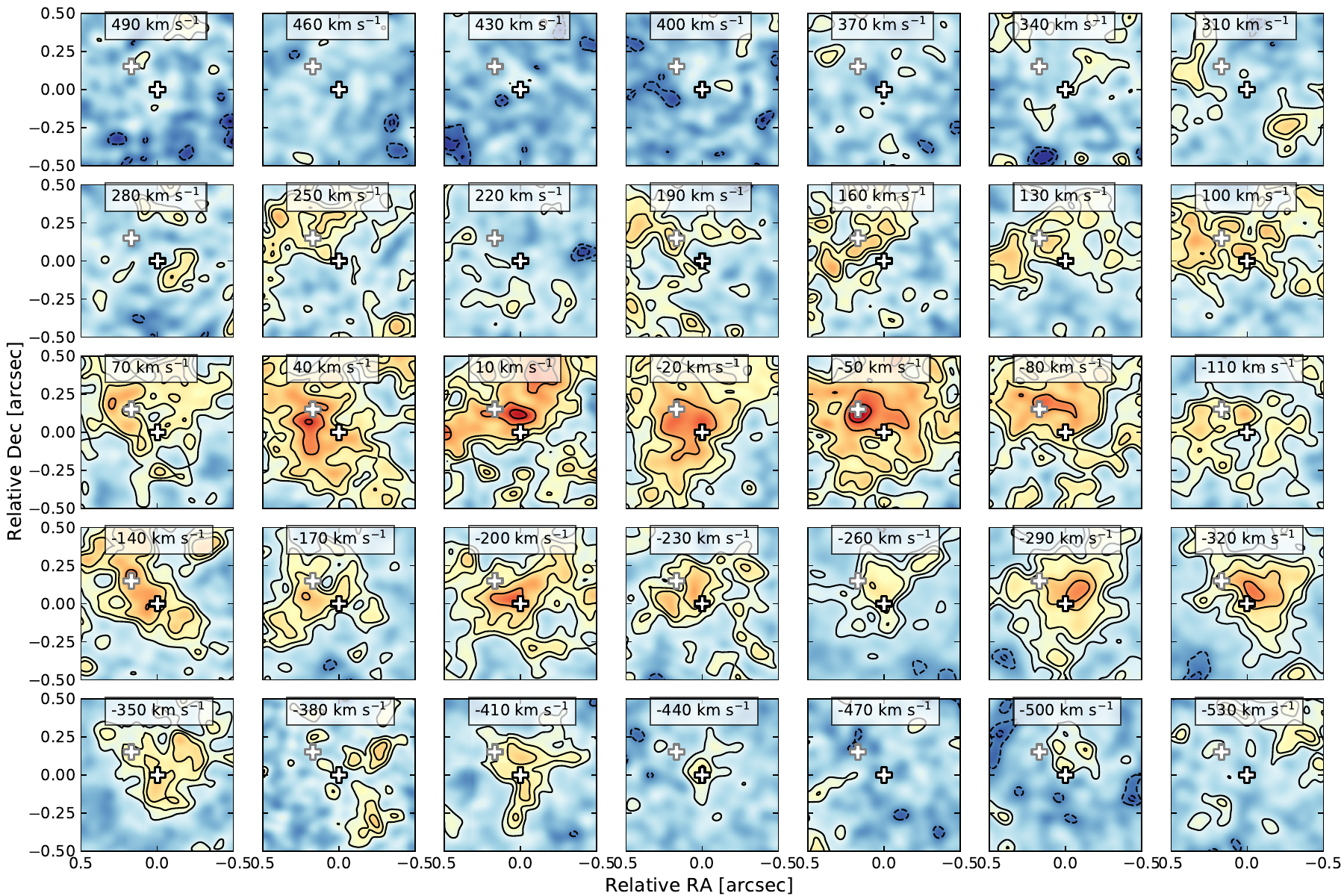}
    \caption{Channel maps of the [CII] emission in the central $1\times1$ arcsec$^2$ region around \zn, in bins of 30 \kms. Contours correspond to [-3,-2,2,3,4,6,8,10]$\sigma$, where $\sigma = 7.8\times10^{-5}$ Jy beam$^{-1}$. The white plus signs mark the quasar position (black contour) and the position of the secondary peak in the continuum emission (gray contour), likely associated with a companion galaxy (Sect. \ref{subsec:alma}). }
    \label{fig:chmaps}
\end{figure*}

Figure~\ref{fig:simbal_ions} shows the continuum normalized spectra of the X-Shooter data and the best-fitting \textit{SimBAL} model.
Both spectra show the difference in the widths of the troughs between HiBALs and LoBALs.
Most notably, \textit{SimBAL} discovered CII BAL that is blended with SiIV trough.
Initial analysis by \citet{Bischetti23} misidentified this feature as part of the SiIV opacity (their Fig.~4) because the overall apparent shape of the blended CII/SiIV trough and CIV trough are similar.
The \textit{SimBAL} analysis revealed the presence of CII BAL that is associated with the AlIII and MgII BALs observed at the low-velocity end.
The ionization potential to create CII ion (11.3 eV) is comparable to those of AlIII (19.9 eV) and MgII (7.6 eV), which suggests that these ions are likely found in a similar region near the hydrogen ionization front in the photoionized gas.
Therefore, it is not surprising that we find observable opacity from CII at the same velocities as AlIII and MgII BALs.

\subsection{Radius of BAL outflow}\label{App:bal_radii}

As discussed in \S~\ref{subsec:simbal}, the gas density ($\log {n}$) cannot be constrained from the spectral analysis due to the lack of absorption lines that are sensitive to the change in $\log {n}$ within the bandpass.
While the outflow properties are tightly coupled to the assumed density of the outflowing gas (Fig.\ref{fig:simbal_radii}), the photoionized gas responsible for the BAL features in \zn\/ can have a wide range of densities and still produce the same absorption features as the strengths of the absorption lines observed in the data do not depend on the density of the gas.
Thus, we make assumptions based on what we know about the radiatively driven outflows and BAL spectral features to provide a plausible range of BAL locations.

The upper limit on the BAL radius ($R_{\rm BAL}=210$ pc) is derived by assuming that the BAL outflow is a steady radiatively driven wind \citep[e.g.,][]{King15,Zubovas&King13}, requiring that the wind kinetic luminosity does not exceed \lbol.
Indeed, if \zn\/ underwent a more luminous quasar (super-Eddington accretion) phase, which is sometimes assumed for high redshift quasars \citep[e.g.,][]{Pezzulli16}, our upper limit may be underestimated.
We derive a lower limit on $R_{\rm BAL}=3$ pc by considering a conservative upper limit on the gas density log($n/\rm cm^{-3}$)$\lesssim 8$, consistent with the highest density values measured in FeLoBAL quasars at $z\lesssim2$ \citep{Choi22}. This can be done by using the best-fit 
 $U$ from \textit{SimBAL} and by applying the definition of ionization parameter,
$U=Q/(4\pi R_{\rm{BAL}}^2nc)$, in which $\log\ Q=57.2$ photons $\mathrm{s^{-1}}$ is the rate of photoionizing photons, estimated from the SED input to {\it Cloudy} scaled to the $L_{3000\mathrm{\AA}}$ measured from the X-Shooter data \citep{Mazzucchelli23} and then integrating for energies greater than 13.6 eV.
A significantly higher density would be similar to that of gas in the quasar broad line region, and the outflowing gas could produce extra emission line features that are not observed in the X-Shooter data (Fig. \ref{fig:simbal_bestfit}), given the high column density. Such a limit on the radius places the BAL gas beyond the broad line region, consistently with our best-fitting \textit{SimBAL} model, where the fractions of partial coverage of the continuum and line emission by BAL outflow are the same. 
Moreover, we did not find evidence for an improvement in the fit when using a different covering factor for continuum and emission lines, in agreement with the BAL gas being located beyond the BLR.
A relevant size scale is the inner edge of the dusty torus, often estimated from the dust sublimation radius. By using equation 
$R_{\rm sub}=0.16(L_{\rm Bol}/10^{45} \mathrm{erg\ s^{-1}})^{1/2}$ pc \citep{Elitzur16} we find  $R_{\rm sub}\sim2.7$ pc, comparable to the lower limit on the BAL wind radius. On the other hand, the upper limit on the radius of the BAL wind is in order of magnitude agreement with the distance from the quasar location of the bulk of the outflowing [CII] gas ($\simeq520$ pc, \S~\ref{subsec:alma}).

\section{Channel maps of the [CII] emission]}

Figure \ref{fig:chmaps} shows the channel maps of the [CII] emission detected in the inner $\simeq5.5$ kpc region around \zn. These maps correspond to 30 \kms\ channels, spanning the velocity range between $\pm500$ \kms, in which most of the [CII] emission is detected (\S~ \ref{subsec:alma}). We find that most of the low-velocity emission is located in the region between the quasar position and the companion galaxy, whose presence is suggested by a second peak in the 242-257 GHz continuum map (Fig. \ref{fig:cii-integrated} left), consistently with a late merger scenario. This is also supported by the fact that we do not see an ordered kinematics in the quasar host-galaxy (Fig. \ref{fig:cii-integrated} right).
On the other hand, we find that the high-velocity, blueshifted ($v\lesssim-250$ \kms) is mostly located close to the quasar position, suggesting the presence of a [CII] outflow in the host-galaxy.

\bibliography{biblio}{}
\bibliographystyle{aasjournal}



\end{CJK}
\end{document}